\documentstyle[preprint,prc,tighten,eqsecnum,aps]{revtex}
\begin{document}
\draft
\preprint {McGill/94--32.}
\title{\bf Fragment flow and the nuclear equation of state}
\author{Jianming Zhang\thanks{email: jzhang@hep.physics.mcgill.ca} and Charles
Gale\thanks{email: gale@hep.physics.mcgill.ca}}
\vspace{1cm}
\address{Physics Department, McGill University \\ 3600 University St.,
Montr\'eal
QC, Canada H3A--2T8}

\maketitle

\begin{abstract}
 We use the Boltzmann-Uehling-Uhlenbeck model with a momentum-dependent nuclear
mean field   to simulate the dynamical
evolution of heavy ion collisions.     We re-examine the azimuthal anisotropy
observable, proposed as sensitive to the equation of state of nuclear matter.
We obtain that this sensitivity is maximal when the azimuthal anisotropy
is calculated for nuclear composite fragments, in agreement with some
previous calculations. As a test case we concentrate on
semi-central $^{197}{\rm Au}\ +\ ^{197}{\rm Au}$ collisions at 400 $A$ MeV.
\end{abstract}
\newpage
\section{Introduction}
It is fair to say that the field of heavy ion
collisions is a flourishing area of contemporary research in physics.
In its higher energy extension, it straggles high energy and nuclear physics.
The main objectives there are to create higher energy densities than ever
attained in terrestrial accelerators before. In this pursuit, one will surely
learn a great deal about the behaviour of strongly interacting matter at high
temperatures and densities.  Ultimately,  one will want to verify
experimentally
one of the most intriguing predictions of QCD: the formation of a plasma of
quarks and gluons, deconfined over macroscopic portions of time and space. A
vigorous experimental program is presently under way and the theoretical
interest, both direct and indirect, being generated is considerable
\cite{qm93}.

Heavy ion collisions in the so-called intermediate energy regime constitute
a unique tool for the investigation of complex nuclear reaction dynamics. As
one goes
beyond the Fermi energy in kinetic energy per projectile nucleon, the phase
space accessible to nucleons in microscopic two-body collisions opens  up
dramatically, owing to the disappearance of Pauli blocking effects. This area
thus offers the intriguing possibility of studying the competition and the
individual effects of the nuclear mean field and two--body collisions. In other
words, the intermediate energy region stretches from a domain
where mean field dynamics
dominate to a regime where microscopic nucleon-nucleon collisions play a major
role.  One of the
main goals of this line of research is an accurate determination
of the bulk properties
of nuclear matter as characterized by the nuclear equation of state (EOS).
The EOS plays a crucial role in the dynamics of heavy ion collisions. It is
also a key ingredient in the theory of stellar collapse leading to
supernov{\ae}
formation. It also naturally  has something to say on neutron star properties.
One realizes the many facets of the nuclear EOS, thumbing through the
proceedings of recent dedicated conferences \cite{penbod}. Information
on the EOS, as characterized generally by the coefficient of
compressibility for nuclear matter in its ground state, K, can also
be deduced from detailed Hartree--Fock plus RPA
analyses of giant monopole resonances in finite nuclei
\cite{blaizot}. These lower energy experiments probe  regions of
excitation energy  adjacent to the nuclear ground state while intermediate
energy heavy ion collisions will create  zones of  high density and
temperature.  Consistency requires that the value of  the nuclear
compressibility coefficient of equilibrium nuclear matter deduced from both
sets of experiments be compatible with one another. Happily, after a period
of apparent disagreement, this goal seems on the verge of being fulfilled.
It now seems that the analysis of giant monopole resonances and of
heavy ion flow data can both  accommodate a value of K $\approx$ 210 MeV
\cite{pear91,jzsdgcg94}. We will  further elaborate on
flow measurements in intermediate
energy heavy ion collisions.

It seems natural that the characteristics of the nuclear EOS would
manifest themselves through some cooperative behaviour like nuclear
collective flow. The experimental identification of this feature was
made possible by the advent of a first generation of 4$\pi$ detectors,
capable of global event reconstruction. Successful schemes used to
plot global variables and quantify this flow were the kinetic flow
tensor  distributions \cite{gus84} and  the average transverse momentum
analysis \cite{do85}.  Some other recent projections of the triple
differential cross section include the azimuthal distributions
\cite{gmw88}and  correlation functions \cite{wan91}.  Calculations  done with
the  very successful BUU model \cite{bdg88} have reported that the azimuthal
anisotropy ratio \cite{gmw88} was an observable sensitive to the value of K
used in the theory. We will address in this paper the issue of the
sensitivity of this particular
observable to the nuclear equation of state.

As the
sophistication in detection techniques increased,
the separate measurement of the flow of nuclear clusters has revealed that
``clusters go with the flow'' \cite{gus88} {\em i.e.} the amount of directed
flow, as characterized by the in-plane transverse momentum per nucleon,  was
found to
{\em increase} with  fragment mass.  This feature had in fact been predicted
rather
early \cite{baum75}.  Other calculations capable of producing nuclear fragments
also contained the feature that flow effects should be stronger, the
heavier the fragment  \cite{sto81,cse83,cse84,pei89}.

One purpose of this paper is to reconcile two apparently contradictory
analyses of neutron azimuthal anisotropy \cite{mad93,elaa94}. We also wish to
provide a quantitative connection between composite flow and the coefficient of
compressibility for equilibrium nuclear matter in the framework of the
BUU model. For the purpose of briefness we will not elaborate on the
well-documented transport model \cite{bdg88} here. We simply give our
arguments in the following sections and we then conclude.

\section{Fragments and the BUU model}
Exactly ten years of calculation with the BUU transport model \cite{bkdg84}
have left no
doubt on the complexity of nucleus-nucleus reactions at intermediate energy and
on the need for a complete transport approach. The model has been quite
successful in reproducing single particle flow patterns and transverse
momentum distributions \cite{gale90,jzsdgcg94}. However, the BUU equation is
the
representation of a one-body theory. It yields the time evolution of the
{\em average} one-body density and consequently it is not well suited to
describe
aspects of nuclear reactions that deal with significant dynamical
branching or fluctuations.
Nuclear multifragmentation is a good example of this class of phenomena. A
significant amount of theoretical activity  has been devoted to incorporate
the effects of fluctuations in the transport approaches. An attempt to extend
the standard BUU in this direction was made by Bauer, Bertsch and Das Gupta
\cite{bbd87}. Also, in an approach similar to the theory of hydrodynamic
fluctuations, a Boltzmann-Langevin equation for the evolution of the one-body
density was used \cite{ag}. A formalism for addressing stochastic one-body
dynamics within the framework of transport theory was devised \cite{ranre}. The
above three techniques have been critically compared in a recent publication
\cite{cha92}. With the exception of the first approach, these scenarios are
still not amenable to calculations that can directly be compared with
experimental results. Some recent important developments involve the use of
quantum many-body theory to derive transport equations with bound-state
production and absorption \cite{danbe91,danpan92}. The work along these
lines seems extremely promising. Note in passing that the
BUU equation without the collision term is the Vlasov equation and the
latter can be obtained from the quantal TDHF equation by taking a Wigner
transform,
then a semiclassical limit \cite{bdg88}. The presence of nuclear clusters
in TDHF final
states has also been indirectly observed \cite{ddg84}.

Driven by the need to interpret the available experimental data in a
plausible fashion, other more phenomenological avenues have been
followed in  the extraction of
composite contributions from transport theory results. One simple and
intuitively appealing approach relies on the idea of coalescence.  This concept
was introduced already long ago \cite{sch63}. The original formulation for
heavy ion collisions was
devised around the thermodynamic model. A
discussion of the coalescence  model and its comparison with other
approaches has appeared
in the literature \cite{jdgm82}. Put simply, the picture stipulates that
if two or more
nucleons are close enough together in phase space when the momentum space
configuration of the reacting system ceases to change, they will emerge as a
self-bound cluster.

In performing theoretical analyses of intermediate
energy heavy ion data and  comparing the results of complete BUU  calculations
with measurements of single
nucleon observables, the need to subtract the ``spurious'' (in this context)
cluster
contribution from the full simulation results has also arisen. Some
early experimental measurements have concentrated on
this independent cluster component. For example, the Plastic Ball group has
observed
relatively large triton yields in nuclear reactions at intermediate and high
energies \cite{doss88}. A coalescence prescription to study the transverse flow
of intermediate mass fragments with a relativistic BUU model has been used
previously and has been shown to provide a good description of the data
\cite{koch90}.  At energies below and around 100 $A$ MeV, a
six-dimensional coalescence model has also been used to filter the
results of VUU simulations and to very successfully compare with
experimental results \cite{kru85}. At such energies,
the composite to free nucleon ratio is larger than at the energies we will
consider here.

It is important to remember that a cluster is really an
entity correlated in six-dimensional phase space. However, in view of the fact
that our BUU approach contains a binding mean field interaction, we shall adopt
a somewhat simpler viewpoint. It is well know from transport theory
calculations that the transverse momentum generation in heavy ion reactions
begins quite early in the history of the reaction and then stops \cite{gale90}.
The
amount of transverse momentum generated has then saturated and the
momentum space distributions are approximately stable. Our idea is to apply a
coalescence criterion in coordinate space only, at this point. Typical BUU
calculations consist of several nucleus-nucleus collisions (``runs'') performed
in parallel to enhance statistics and  to provide a smooth initial state
density profile in
coordinate and momentum space \cite{bdg88}\footnote{An alternate viewpoint is
that each ``physical'' nucleon is represented by a number of ``test nucleons''
equal to the number of BUU runs.}.  The approach is then the
following:  within a given BUU run, a nucleon will be considered
``free'' only if no other nucleons
are found within a certain critical three-dimensional distance, $d_c$.
Otherwise, it will be
considered a component of a bound cluster. We justify restricting our analysis
to
coordinate space by the fact that, owing to the dynamical nature of the
problem and to
single particle propagation in the transport model, particles nearby in
coordinate space but far apart in
momentum space will separate after a certain time. There
are two parameters to our scenario: the time at which the coalescence
model is applied, $t_c$,  and the critical distance parameter, $d_c$. We
choose $t_c$ as the time in the nucleus-nucleus centre of mass frame when
the transverse momentum generation just starts to
saturate. This is calculated for each reaction we study. The value of $d_c$ is
left as a free parameter and adjusted to experimental data (see the next
section).  The  coordinate space coalescence + BUU procedure is
in fact not new \cite{aibe85} and  has
recently been applied to semi-central Au + Au reactions at 150, 250, 400 and
650 $A$ MeV \cite{wmzhang94}. At the two lower beam energies, the BUU
calculations with the simple coalescence prescription overestimate somewhat the
triple differential cross sections for free neutrons with laboratory
polar angles above 15$^\circ$. They however agree well with
the neutron data from collisions at 400
and 650 $A$ MeV  over a wide range of laboratory polar angles. This approach
has
also successfully treated the case of Nb + Nb at 400 $A$ MeV
\cite{elaa94,elaa94b}.

\section{Collective flow }

There are several experimental observables that have been proposed as
a quantitative measure of the collective flow in heavy ion collisions. Here we
shall mainly concentrate on azimuthal distributions and on the so-called
maximum
azimuthal anisotropy ratio \cite{gmw88}. Those azimuthal distributions are
measured event-by-event with respect to the reaction plane which can be
estimated
experimentally. There is always some error associated with the
reaction plane determination
\cite{dan88}.  The maximum global azimuthal anisotropy ration can be defined as
\begin{equation}
{\cal{R}}\ =\
\frac{\frac{d\sigma}{d\phi}|_{\phi=0^\circ}}{\frac{d\sigma}
{d\phi}|_{\phi=180^\circ}}\ .
\end{equation}
Microscopic BUU calculations have shown that the maximum
azimuthal anisotropy ratio for all test nucleons in a chosen
rapidity range, $\cal{R}$,
was sensitive to the value of the
compressibility coefficient for equilibrium nuclear matter used in the theory
\cite{gmw88}. A maximum azimuthal anisotropy
ratio can also be defined at each polar angle, in a given rapidity range:
\begin{equation}
r(\theta)=\frac{\sigma_{3}(\theta,\phi)|_{\phi=0^{o}}}
{\sigma_{3}(\theta,\phi)|_{\phi=180^{o}}}\ ,
\label{eq:rr}
\end{equation}
where
\begin{equation}
\sigma_{3}(\theta,\phi)\equiv d^{3}\sigma/d(cos\theta)d(\phi-\phi_{R})dy \ .
\label{eq:ss}
\end{equation}

The variables defined above have been determined in  recent experiments
measuring
triple-differential cross sections of neutrons emitted in semi-central
heavy ion collisions of Nb on Nb at 400 $A$ MeV.  The data
were confronted with
BUU calculations with the simple coalescence prescription to single-out the
contribution of free neutrons \cite{elaa94,elaa94b}. This turned out to
be a rather necessary and successful ingredient. With
this prescription, the polar-angle
dependence of the maximum azimuthal anisotropy ratio $r ( \theta ) $ for
the measured neutrons,  emitted with rapidity 0.7 $\leq \
(y/y_{\rm beam})_{\rm c.m.}\ \leq$ 1.2, could be reproduced by the
calculations. We could also reproduce the
triple-differential cross section for emitted neutrons both in magnitude and
behaviour  \cite{elaa94,elaa94b}. The coalescence parameter $t_c$ was
determined
according to the criterion introduced in the previous section. The parameter
$d_c$ was adjusted such that the model would reproduce the double-differential
cross section for emitted neutrons. This single parameter fit reproduced well
the double differential cross section in magnitude and polar angle
dependence. We have imposed a further check of our simple
coalescence picture by scanning each identified cluster and keeping only those
nucleons that could belong kinematically to a common Fermi sphere in momentum
space. The change in our results was less than 1\%.
One conclusion of the experimental investigation and its comparison with theory
was  that the maximal azimuthal anisotropy
ratio $r ( \theta ) $ of {\em free} neutrons turned out {\em not} to be
sensitive to the nuclear EOS contrary to what was hoped previously. However,
full one-body calculations of $r ( \theta ) $ do exhibit considerable structure
and sensitivity to K \cite{mad93}, as claimed originally \cite{gmw88}. They
also overpredict the triple differential
neutron cross sections.

For the purpose of clarifying the behaviour of the above observables
in the theory,
we concentrate here on the case of semi-central collisions of Au + Au at 400
$A$
MeV. The impact parameter range we shall integrate over is $0\ \leq\ {\rm b
(fm)}\ \leq\ 6.2$, and $d_c$ was set at 3.2 fm. These conditions were
determined in
a recent
investigation of free neutrons emitted in heavy ion collisions
\cite{wmzhang94}. The compressibility coefficient, K,  for equilibrium nuclear
matter  can be chosen by varying the set of constrained parameters
A, B, C, $\sigma$ and $\Lambda$ in the MDYI momentum-dependent nuclear mean
field \cite{gmw88}:
\begin{equation}
U(\rho,\vec{p})=A[\frac{\rho}{\rho_{0}}]+B[\frac{\rho}{\rho_{0}}]^{\sigma}
+2\frac{C}{\rho_{0}}\int\!\,d^{3}p^{'} {f(r, \vec{p}^{\ '}) \over
1+[\frac{\vec{p}-\vec{p}{\ '}}{\Lambda}]^{2}}\ .
\label{eq:u1}
\end{equation}
The parameters used in this work will correspond to K = 100, 150, 215, 250, and
380
MeV. Our calculations include Coulomb effects.
Fig. {\ref{fig1}}a shows the maximum azimuthal anisotropy ratio
$r(\theta)$ of
free nucleons plotted against
laboratory polar angle $\theta$ for near-central Au + Au
collisions at 400 $A$ MeV. For
now, we
restrict our analysis to the rapidity region 0.7 $\leq\ (y/y_{\rm beam})_{\rm
c.m.}\ \leq$ 1.2 \cite{elaa94,wmzhang94,elaa94b}. Furthermore we
implement a hard ``spectator
cut'', requiring that the momentum of the BUU test particles be larger than
0.25 GeV/c in the projectile rest frame and the laboratory frame.
These cuts
essentially remove particles emitted with $\theta \alt 12^{\circ}$ and
$\theta \agt 30^{\circ}$.
The statistical uncertainties in the calculation will be slightly larger near
the edges of the populated region.   As claimed in Ref.~\cite{elaa94,elaa94b},
these
results further confirm that the free nucleon azimuthal data are
essentially insensitive to variations in the nuclear EOS.  In Fig.
{\ref{fig1}b}, we show $r(\theta)$ {\em vs.} $\theta$ for all BUU nucleon test
particles.  A much clearer
sensitivity to the nuclear compressibility K can now be seen over most of the
covered polar angle range. We may now subtract the free nucleons identified
with our
coalescence prescription to obtain a signal due to all the clusters
averaged-over in the one-body BUU. This is show in Fig. {\ref{fig1}c}. Clearly,
the
highest values of the azimuthal anisotropy ratio are reached with the clusters
only. A strong variation with K is also observed.    In Fig. \ref{fig2} we plot
the polar angle integrated azimuthal distributions for free nucleons and
clusters, using the procedure described above. Both distributions peak at
$\phi = 0^{\circ}$.
The azimuthal differential cross
sections for composites are considerable larger than
those for free nucleons for small azimuthal angles $\phi$ and become
comparable at large $\phi$. The width of the composite distribution
is also significantly smaller. Thus our calculations are entirely consistent
with
the experimental observation that fragment flow is more
correlated with the reaction plane than that of single particles \cite{doss87}.

We now
calculate the maximum global anisotropy ratio $\cal{R}$, subject to the same
kinematical rapidity and spectator cuts as before.
We plot $\cal{R}$ against the coefficient
of compressibility for equilibrium nuclear matter, K, in Fig. \ref{fig3}.
Each point in this figure represents a set of impact parameter integrated BUU
calculations.
The power and simplicity
of this plot \cite{mad93} is immediately apparent: an experiment
measures one value of
$\cal{R}$, given a well defined set of kinematical constraints.   This would
appear on this plot as an horizontal line. The intercept of this line
with the appropriate theoretical curve would then directly yield a
value of K. The free nucleons
do not constitute a very sensitive observable, as we
can see: the
steeper the curve, the more accurate is the deduced value of K. All the test
nucleons analyzed together are somewhat sensitive to the value of K, but the
clusters
alone are much more sensitive. As in the past one-body theories such as the BUU
have
compared with ``pseudo-nucleons'' obtained from folding all the measured
particles (free nucleons and composites) together \cite{gos89,aic89},
an analogous procedure can be followed to produce a
value of $\cal{R}$ to be interpreted with the top curve in our Fig. \ref{fig3}.
The values of $\cal{R}$ obtained here with the full BUU test particle ensemble
are comparable
in magnitude with those of the original study \cite{gmw88}, done with a
different system at a slightly different bombarding energy.

The usefulness of the  azimuthal anisotropy ratio depends largely on an
accurate determination of the event reaction plane. A method which
circumvents this difficulty is based on the azimuthal pair correlation function
\cite{wan91}:
\begin{equation}
C(\psi) =\frac{P_{cor}(\psi)}{P_{uncor}(\psi)}\ ,
\end{equation}
where $\psi$ is the azimuthal angle between the transverse momenta of two
particles. $P_{cor}(\psi)$ is the $\psi$ distribution for
observed pairs from the
same event, and $P_{uncor}(\psi)$ is the $\psi$ distribution for pairs
from mixed events. Following Ref. \cite{gmw88}, let us assume that the
azimuthal cross section has the form
\begin{equation}
\frac{d\sigma}{d(\phi-\phi_{R})dY} = a[1+\lambda cos(\phi-\phi_{R})]
\label{azcr}
\end{equation}
where $\phi_{R}$ is the azimuthal angle of the reaction plane. If this
parametrization is exact, $C(\psi)$ can be written as
\begin{equation}
C(\psi) = 1+\frac{1}{2}\lambda^{2}cos (\psi )\ .
\label{cpsi}
\end{equation}
The maximum global azimuthal anisotropy ratio defined earlier can be
expressed as
\begin{equation}
{\cal{R}} =
\frac{\frac{d\sigma}{d\phi}|_{\phi=0^\circ}}{\frac{d\sigma}{d\phi}|_{\phi=
180^\circ}}=\frac{1+\lambda}{1-\lambda}\ .
\end{equation}
Thus, measuring the azimuthal pair correlation function can immediately provide
us with the anisotropy ratio, $\cal{R}$, without the ambiguities associated
with event-by-event reaction plane determination. We can easily verify the
validity of Eq. \ref{azcr} as an accurate parametrization in the
case at hand. Since
the issue of statistics is quite important here, we shall slightly shift our
rapidity window. We shall choose $0.4\leq (y/y_{\rm proj})_{\rm cm}\leq
0.8$ as a sufficiently
populated region with a reasonable sensitivity to the nuclear EOS, other
kinematical constraints being the same.  We study the
same reaction as before, with the same reaction parameters. Fig. {\ref{fig4}a}
is a plot of the azimuthal cross section for all test particles, calculated
with three different equations of state: K = 100, 215, and 380 MeV. The curves
represent a fit with the parametrization of Eq. \ref{azcr}. Fig. {\ref{fig4}b}
represents the same exercise repeated for clusters as defined through our
coalescence
prescription. In both cases the accuracy of the cosine assumption is
remarkable. We plot on Figs. {\ref{fig5}a} and {\ref{fig5}b} the azimuthal pair
correlation function as calculated numerically together with the
parametrization, Eq. \ref{cpsi}. The value of $\lambda$ used are the same ones
obtained by fitting the results in Fig. \ref{fig4}. Again, the agreement
between the analytic formul{\ae} and the numerical results is excellent. We can
now easily plot the azimuthal anisotropy ratio as a function of compressibility
as before. See Fig. \ref{fig6}. The results are very
similar to those obtained with the rapidity
window 0.7 $\leq\ (y/y_{\rm beam})_{\rm c.m.}\ \leq$ 1.2, but with somewhat
smaller numerical values. In cases where the cosine form is not appropriate or
when the reaction plane is simply not measured, one can define a slightly
modified
definition of a global  azimuthal anisotropy ratio:
\begin{equation}
{\cal{R}}^\prime\ =\ \frac {C ( \psi = 0^\circ )}{C ( \psi = 180^\circ )}\ .
\end{equation}
In the theory, this quantity has the same desirable
behaviour with respect to variations of the
nuclear matter compressibility coefficient as $\cal{R}$ as seen in Fig.
\ref{fig7}. It can   then also be
useful in comparisons of experimental results with theoretical calculations.

\section{Summary}
In the framework of the BUU model, using a simple coalescence prescription we
have provided an explanation for an apparent discrepancy between
two calculations of
the azimuthal anisotropy ratio in intermediate energy heavy ion
collisions. We believe that it is quite important from the point of view
of consistency to resolve this issue within the framework of the very
model used to propose this experimental observable.
This involved emphasizing the role played by the nuclear composite fragments
and
their participation in the collective nuclear flow. We have found behaviours
in qualitative agreement with earlier calculations with different models and
with existing experimental data. It is
clear that our approach to clustering in heavy ion reactions  is simple
but we believe that more sophisticated
scenarios should reach similar physical conclusions. What is also
evident is that the aim of the theoretical efforts in
understanding the dynamics of heavy ion collisions should not be
restricted to the mere extraction of a single number, K. High energy nuclear
collisions clearly constitute a rich
problem, with many possible projections of the experimental data to be dealt
with.

Again, our aim was to extract some physics content in an admittedly
phenomenological fashion, not to provide a rigorous and complete theory of
cluster production. The theoretical problems associated with a complete
time-dependent theory of
composite formation are being addressed. We believe that therein lies the key
to a more complete understanding of the nuclear dynamics involved in heavy ion
collisions. In this respect, it is quite stimulating that very high
quality data is becoming
available. For example the EOS TPC collaboration has started to release some
experimental results
\cite{tpc}. This data is  virtually free of
experimental biases. The EOS Time Projection Chamber, with its simple and
seamless acceptance, good particle identification and high statistics, was
designed to overcome some of the limitations of the previous generation of
4$\pi$
detectors. Composites up to high $Z$ have been measured in the TPC.
This fact,
combined with the active   program under way at the GSI/SIS facility enables
us to look towards the future with optimism.

\acknowledgments

It is a pleasure  to acknowledge  discussions with S. Das Gupta,  D.
Keane and  R. Madey.
Our research is supported in part by the Natural
Sciences and Engineering Research Council of Canada and in part by the
FCAR fund
of the Qu\'ebec Government.

\clearpage


\clearpage
\begin{figure}
\caption{ {\bf (a)} We plot the azimuthal anisotropy ratio for free nucleons at
a
given polar angle, as a
function of $\theta_{\rm lab.}$. The different symbols correspond to different
values of the compressibility coefficient of equilibrium nuclear matter, in
units
of MeV. The reaction under scrutiny is $^{197}$Au + $^{197}$Au at 400 $A$ MeV.
The multiplicity cut corresponds to semi-central collisions. Other kinematical
constraints are specified in the main text. The statistical uncertainties are
of the order of 10\% in the middle of the populated area in polar angle and
roughly 20\% at the edges.  {\bf  (b)} Same caption as (a),
except that all BUU test particles satisfying the kinematical cuts are
involved. {\bf (c)} Same caption as (a),
except that all free nucleons have been subtracted from the full set of BUU
test particles satisfying the kinematical cuts.}

\label{fig1}
\end{figure}
\begin{figure}
\caption{{\bf (a)} The azimuthal distributions of free nucleons with respect to
the
reaction plane are
plotted. The calculations shown correspond to three different values of K. The
units of K are MeV. The curves were drawn to guide the eye. {\bf (b)} Same
caption as (a) but for nuclear clusters.}
\label{fig2}
\end{figure}
\begin{figure}
\caption{We plot the maximum global azimuthal anisotropy ratio as
defined in the main text, as a function of the compressibility coefficient for
equilibrium nuclear matter. The kinematical  cuts are such that particles with
rapidity $y$ such that 0.7 $\leq \
(y/y_{\rm beam})_{\rm c.m.}\ \leq$ 1.2 were accepted. The spectator cut as
defined in the main text was also implemented. The curves were drawn to guide
the eye.   }
\label{fig3}
\end{figure}
\begin{figure}
\caption{{\bf (a)} The azimuthal distributions of free nucleons with respect to
the
reaction plane are
plotted. The calculations were repeated from three different values of K. The
units of K are MeV. The rapidity window has been shifted to 0.4 $\leq \
(y/y_{\rm beam})_{\rm c.m.}\ \leq$ 0.8. The curves are not drawn through the
data but represent a fit with Eq. (3.6). {\bf (b)}  Same caption as (a) but for
nuclear clusters.}
\label{fig4}
\end{figure}
\begin{figure}
\caption{{\bf (a)}The azimuthal correlation function, calculated with all the
BUU test
nucleons is plotted. The curves represent Eq. (3.7) with values of $\lambda$
obtained from Fig. 4. {\bf (b)}  Same caption as (a) but for
nuclear clusters. }
\label{fig5}
\end{figure}
\begin{figure}
\caption{We plot the maximum global azimuthal anisotropy ratio as
defined in the main text, as a function of the compressibility coefficient for
equilibrium nuclear matter. The kinematical  cuts are such that particles with
rapidity $y$ such that 0.4 $\leq \
(y/y_{\rm beam})_{\rm c.m.}\ \leq$ 0.8 were accepted. The spectator cut as
defined in the main text was also implemented. The curves were drawn to guide
the eye. }
\label{fig6}
\end{figure}
\begin{figure}
\caption{We plot the modified global azimuthal anisotropy ratio as
defined in the main text, as a function of the compressibility coefficient for
equilibrium nuclear matter. The kinematical  cuts are such that particles with
rapidity $y$ such that 0.4 $\leq \
(y/y_{\rm beam})_{\rm c.m.}\ \leq$ 0.8 were accepted. The spectator cut as
defined in the main text was also implemented. The curves were drawn to guide
the eye.}
\label{fig7}
\end{figure}
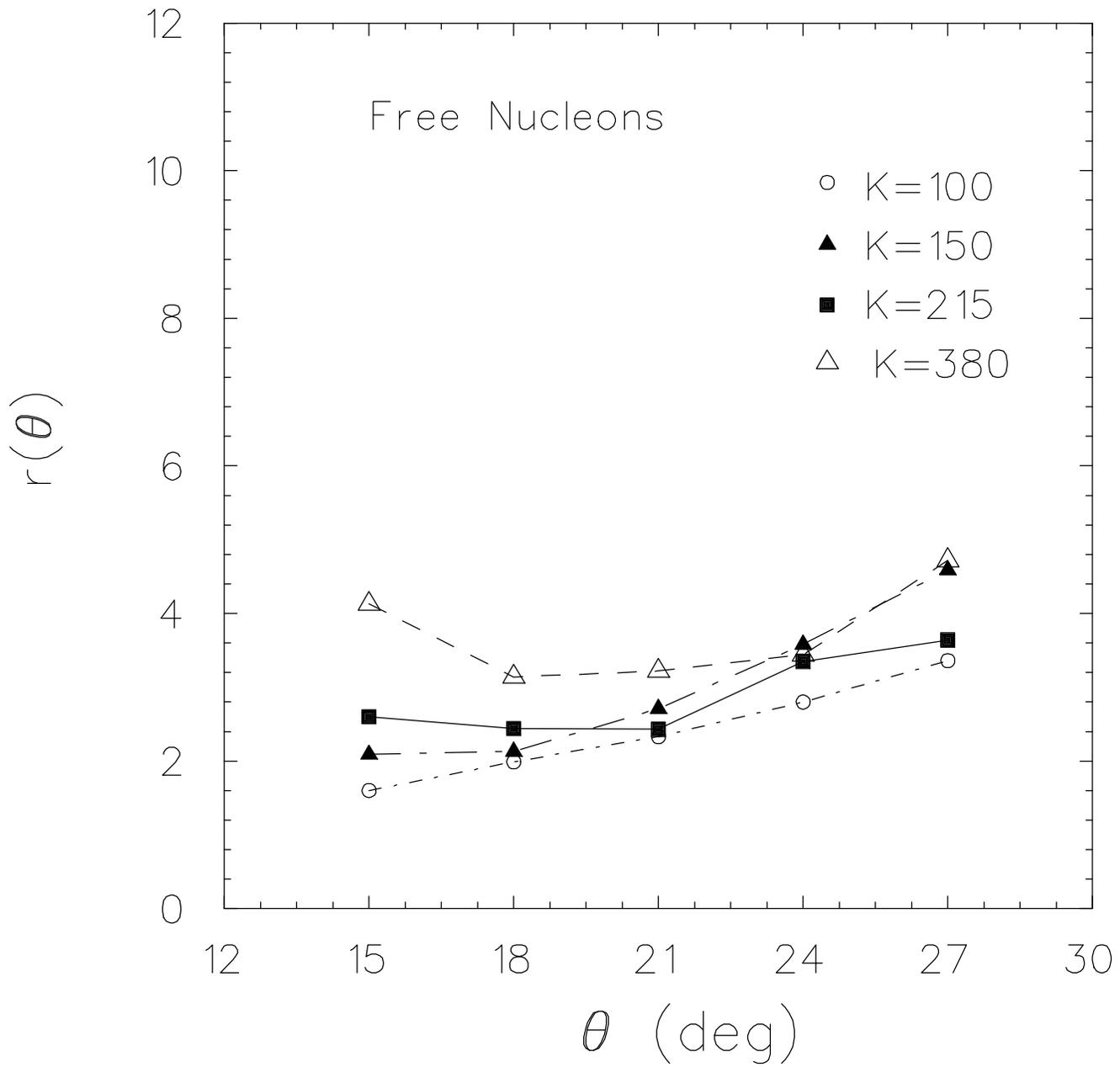
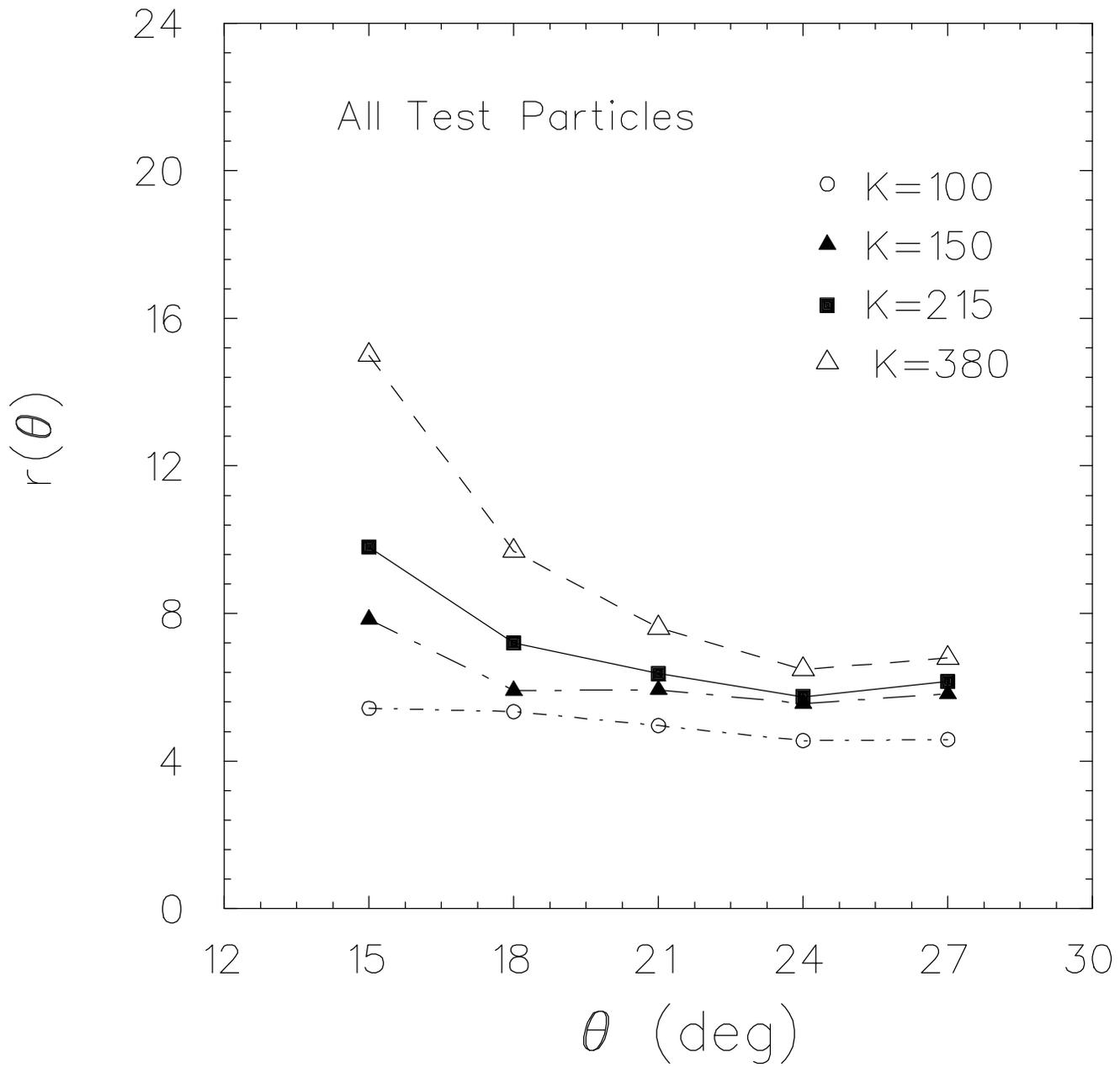
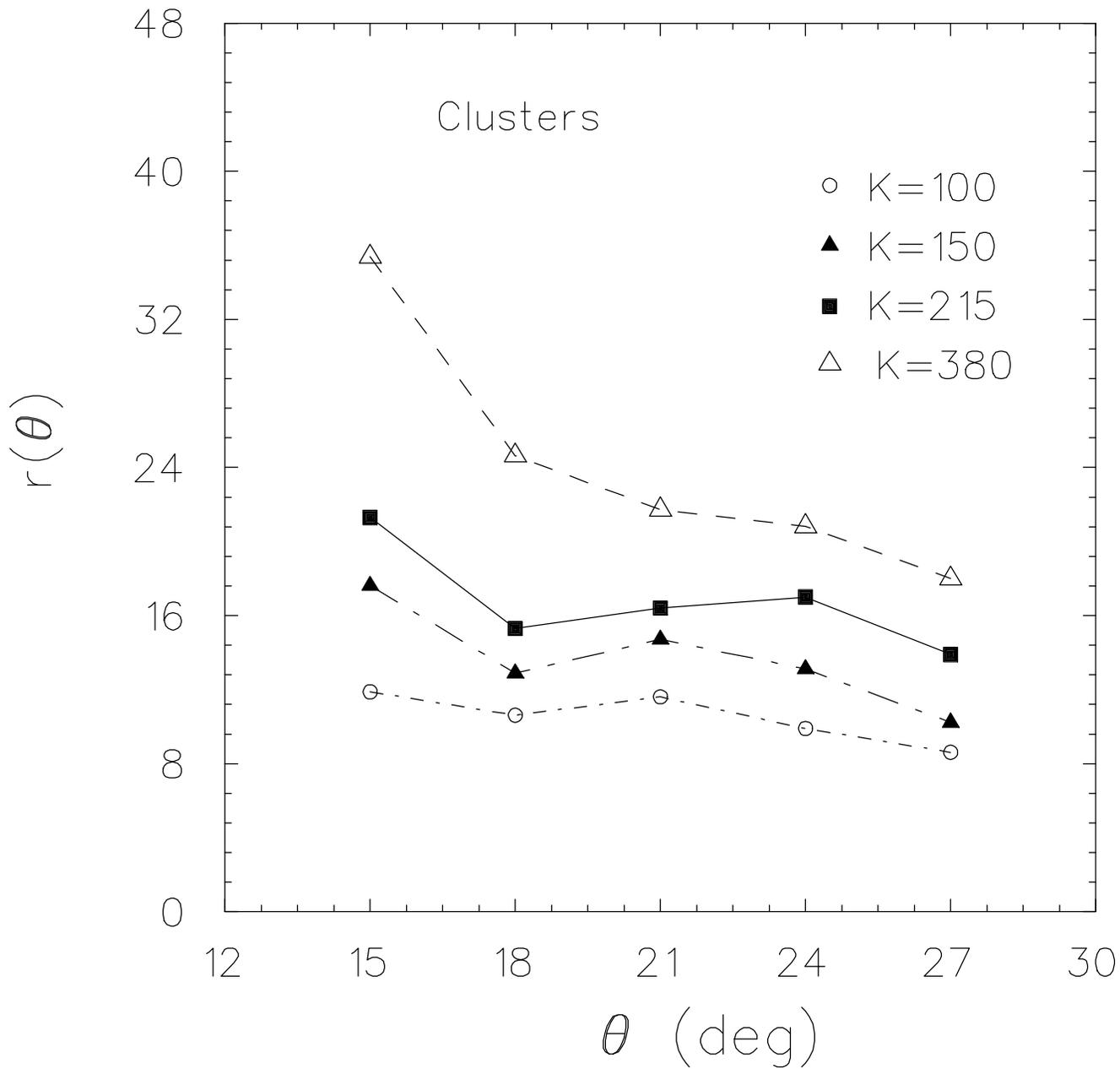
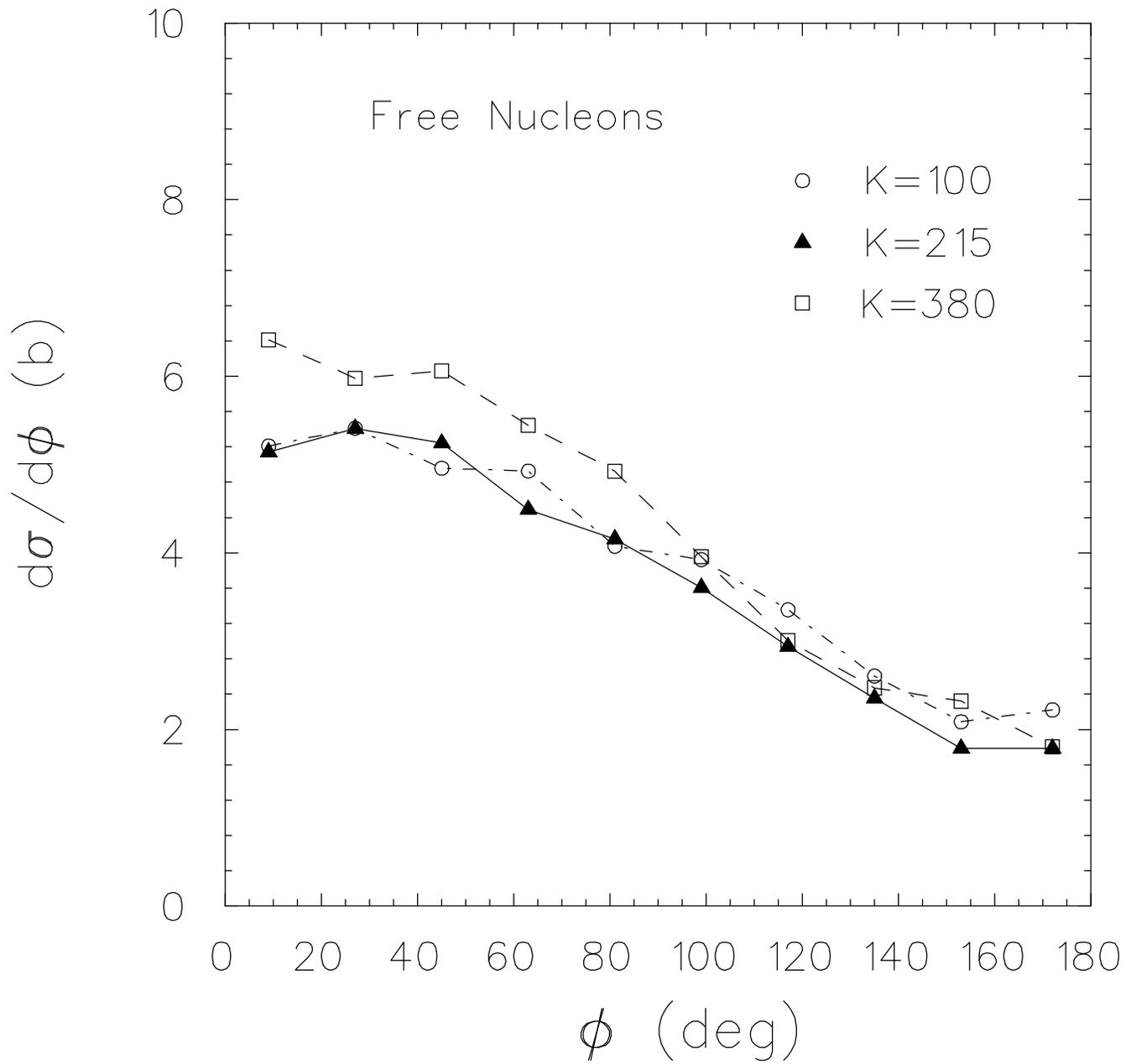
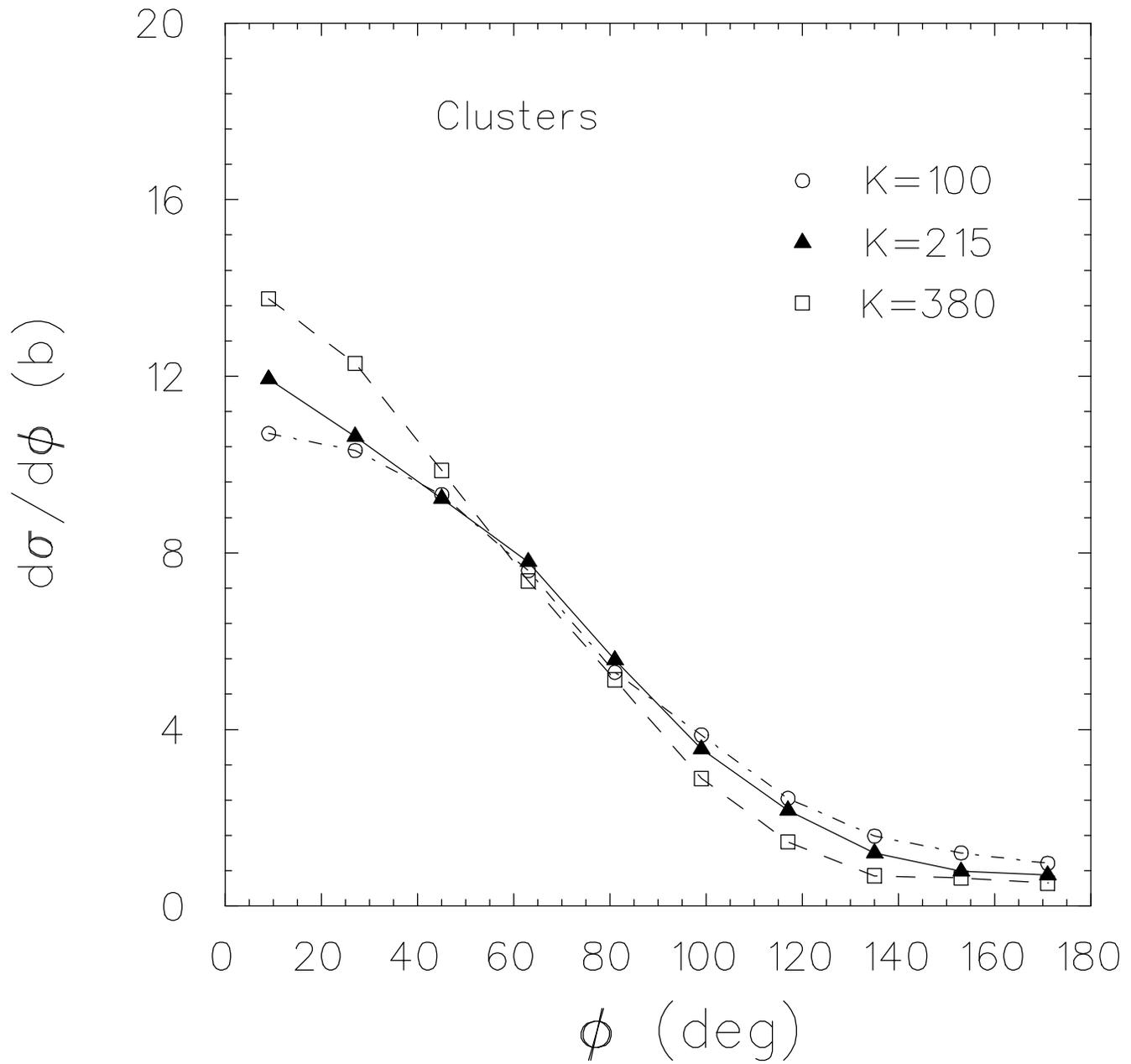
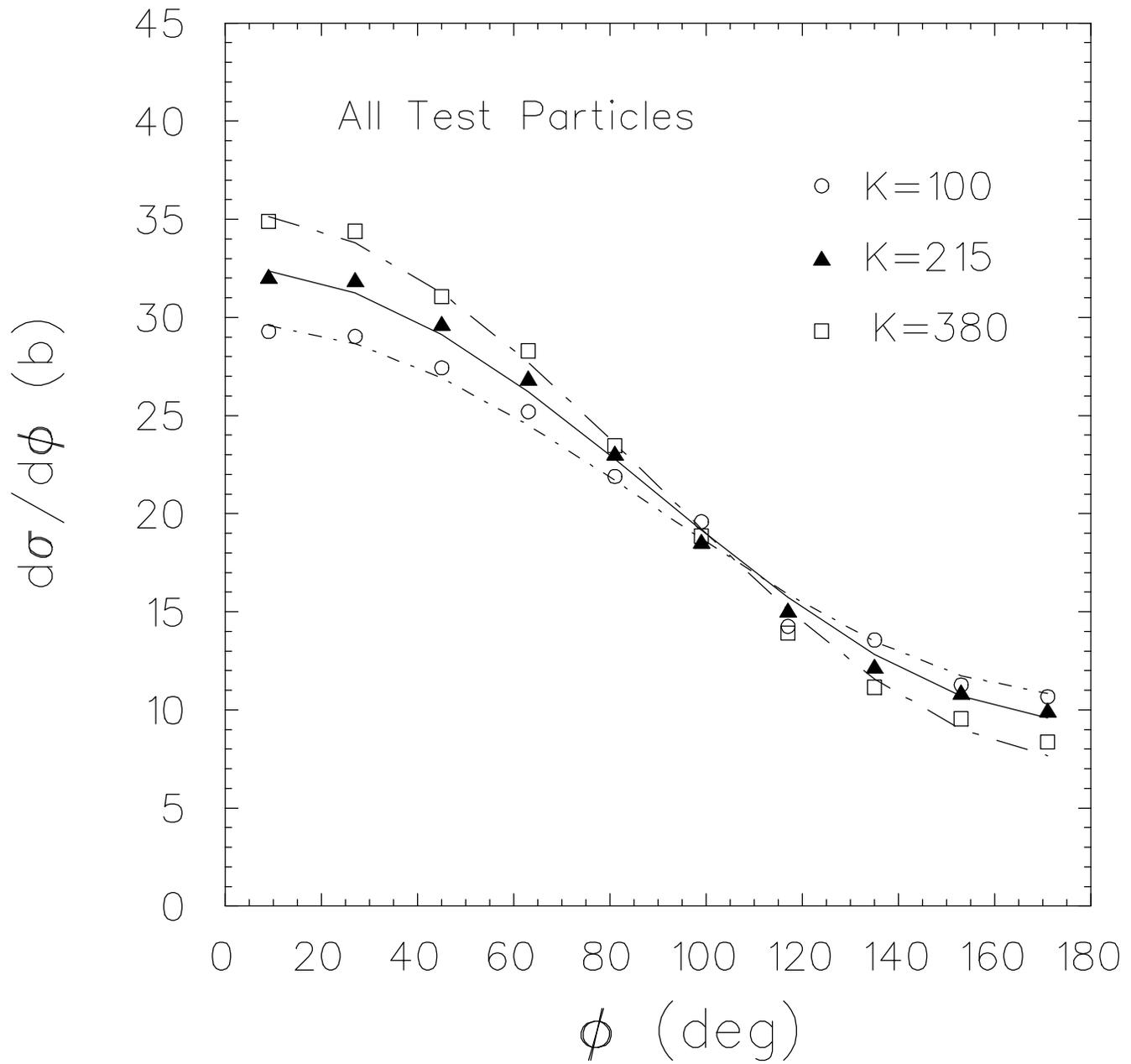
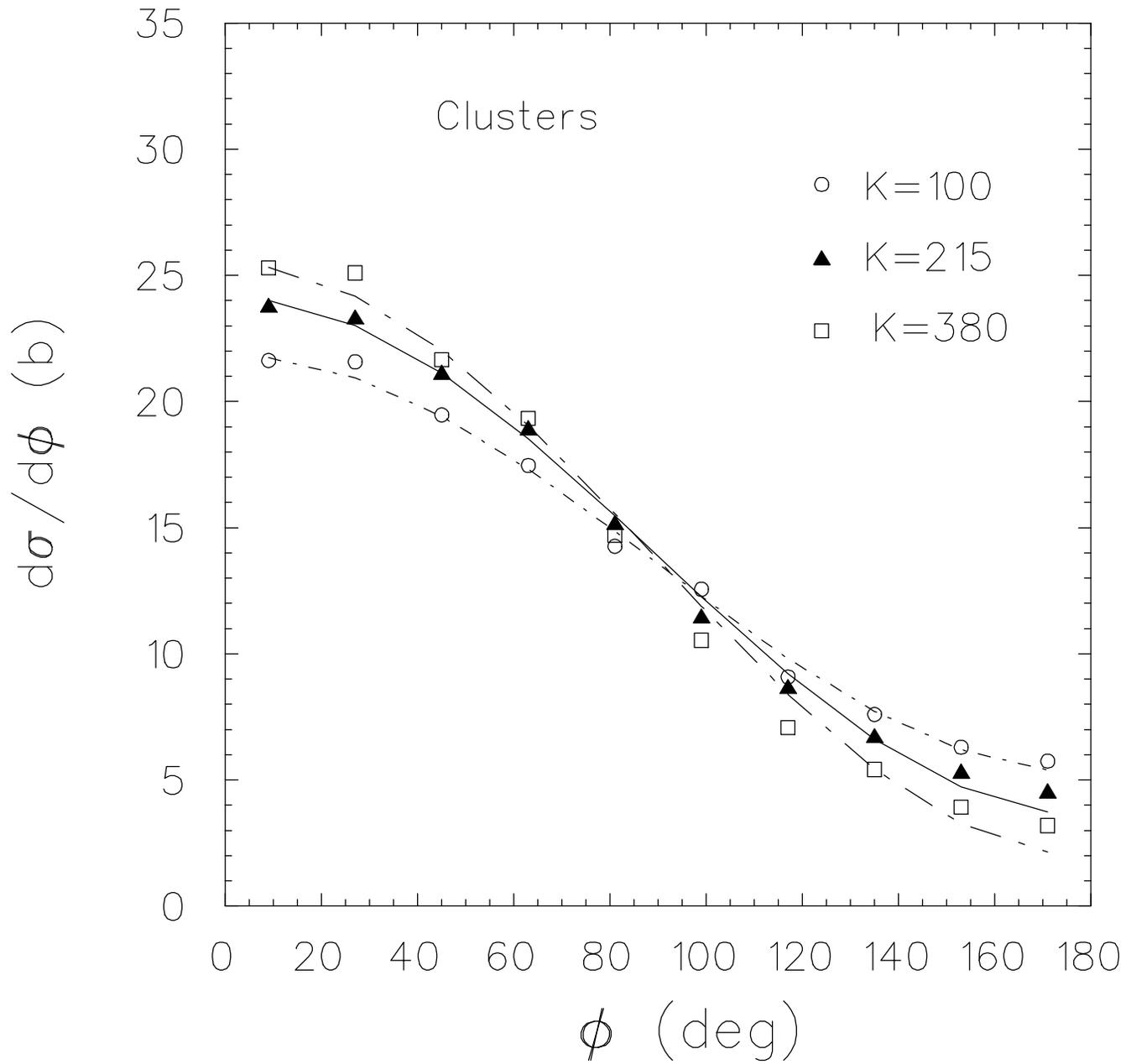
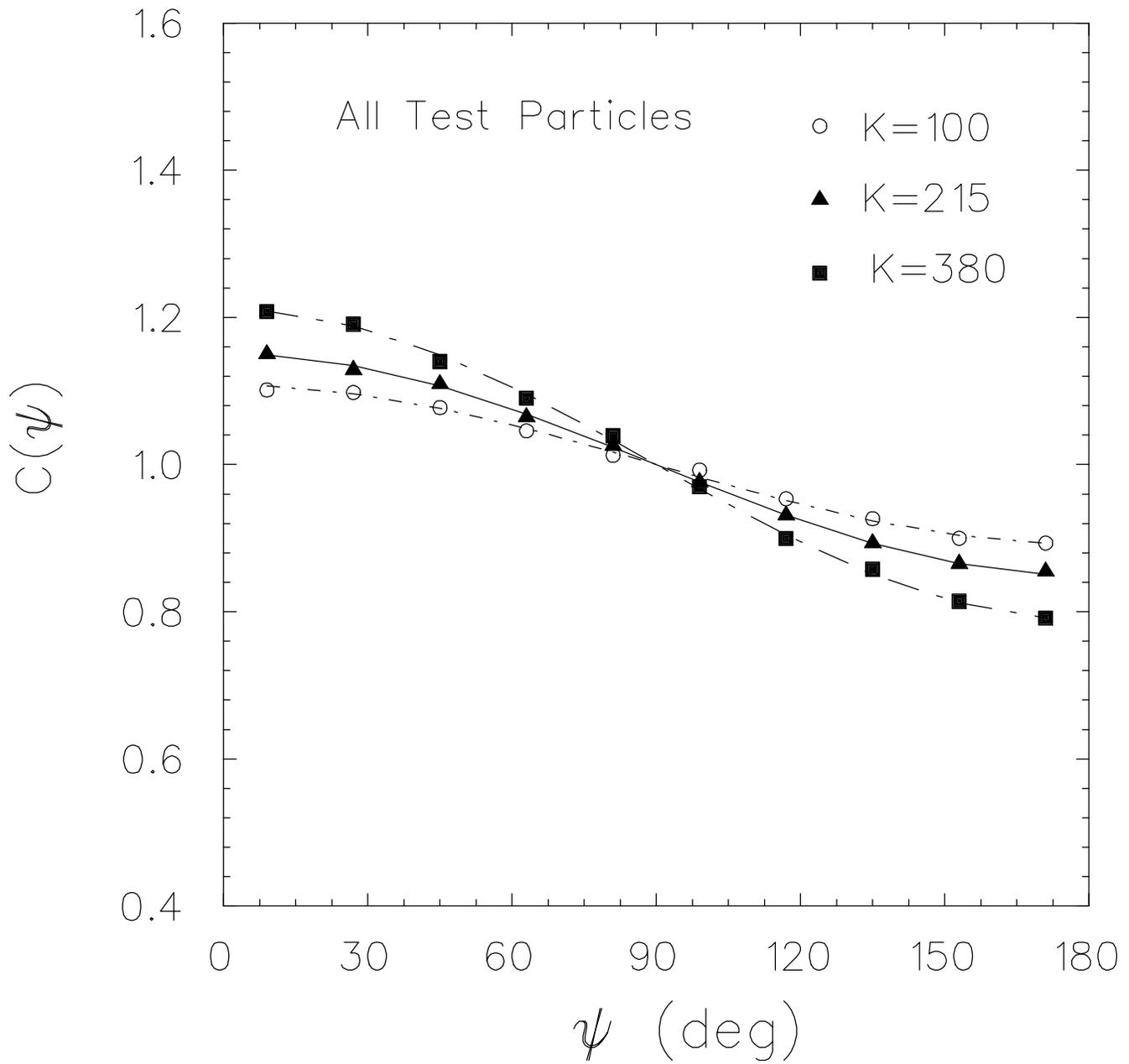
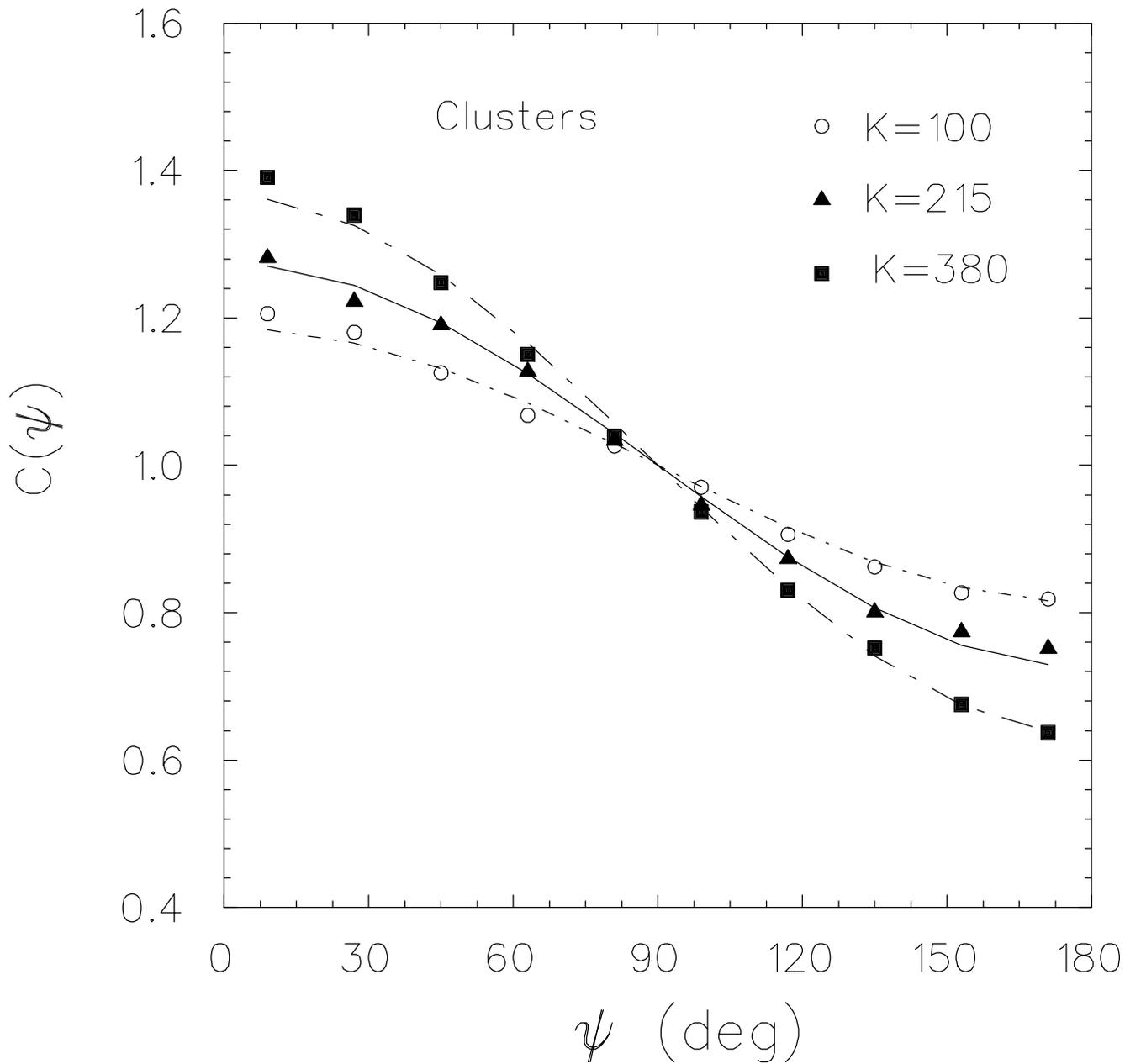
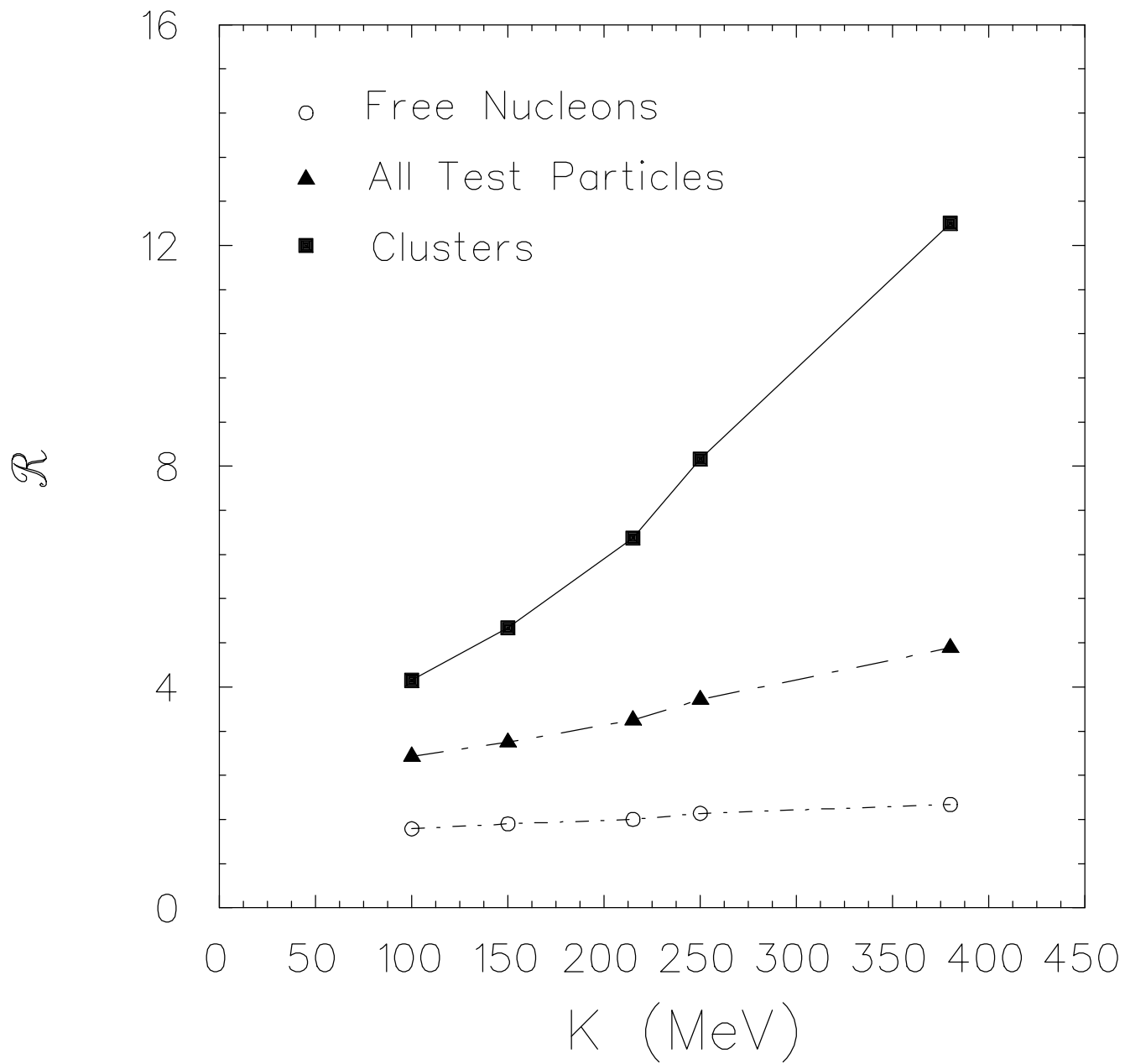
\end{document}